\documentclass{elsart3p}

\usepackage{graphicx}
\usepackage{graphics}
\usepackage{amssymb,amsmath}
\usepackage{bm}

\begin{document}

\begin{frontmatter}



\title{Heat transport and electron cooling in ballistic normal-metal/spin-filter/superconductor junctions}


\author[address1]{Shiro Kawabata},
\author[address2]{Andrey S. Vasenko},
\author[address3]{Asier Ozaeta},
\author[address3,address4]{Sebastian F. Bergeret},
\author[address2]{Frank W. J. Hekking}

\address[address1]{Electronics and Photonics Research Institute (ESPRIT), National Institute of Advanced Industrial Science and Technology (AIST),  Tsukuba, Ibaraki, 305-8568, Japan}
\address[address2]{LPMMC, Universit\'{e} Joseph Fourier and CNRS, 25 Avenue des Martyrs, BP 166, 38042 Grenoble, France}
\address[address3]{Centro de F\'{i}sica de Materiales (CFM-MPC), Centro Mixto CSIC-UPV/EHU, Manuel de Lardizabal 5, E-20018 San Sebasti\'{a}n, Spain}
\address[address4]{Donostia International Physics Center (DIPC), Manuel de Lardizabal 5, E-20018 San Sebasti\'{a}n, Spain}
%\corauth[cor1]{Corresponding author. Tel.: +81-29-861-XXXX; fax: +81-29-861-XXXX, {\it E-mail address}: s-kawabata@aist.go.jp}

\begin{abstract}
We investigate electron cooling based on a clean normal-metal/spin-filter/superconductor junction.
Due to the suppression of the Andreev reflection by the spin-filter effect, the cooling power of the system is found to be extremely higher than that for conventional normal-metal/nonmagnetic-insulator/superconductor coolers.
Therefore we can extract large amount of heat from normal metals.
Our results strongly indicate the practical usefulness of the spin-filter effect for cooling detectors, sensors, and quantum bits.
\end{abstract}

\end{frontmatter}

% main text
\newpage

\section{Introduction}
The quasiparticle transport across a normal-metal/insulator/superconductor (N/I/S) junction is governed by single and Andreev processes.
When the energy $E$ of quasiparticles is larger than the superconducting gap $\Delta$, single quasiparticles can tunnel through the barrier I.
This selective tunneling of "hot" quasiparticles gives rise to electron cooling of the normal metal in an N/I/S junction~\cite{rf:Giazotto06,rf:Virtanen,rf:Muhonen}.
Experimentally, the cooling of a normal metal from 300mK down to below 100mK has been demonstrated~\cite{rf:Giazotto06,rf:Nahum}.

On the other hand, an energy $E$ below the gap ($E < \Delta$), as a result of the Andreev reflection, two quasiparticles can tunnel into S from N and form a Cooper pair in the S electrodes. 
A limitation of the performance of N/I/S coolers is resulting from the such two-particle Andreev processes.
The Andreev current does not transfer heat through the N/I/S interface but rather generates so called the Andreev Joule heating~\cite{rf:Bardas,rf:Rajauria,rf:Vasenko}.
At low temperature regimes, the Andreev Joule heating exceeds the single-particle cooling.

A simple way to enhance the cooling power is to reduce the N/I/S junction transparency. 
However, small barrier transparency hinders "hot" single-quasiparticle transport and leads to a serious limitation in the achievable cooling powers.
In order to increase the barrier transparency and to reduce the Andreev Joule heating, it was suggested to use ferromagnetic metals (FM) as an interlayer~\cite{rf:Giazotto02,rf:Burmistrova,rf:Ozaeta}.
Giazotto and co-workers have investigated the cooling of a clean N/FM/S junction theoretically and found the enhancement of the cooling power compared to conventional N/I/S junctions due to the suppression of the Andreev Joule heating~\cite{rf:Giazotto02}.
However in order to realize such an efficient cooler, impractical FMs with extremely-high spin-polarization $P > 0.94$ like half metals~\cite{rf:Wolf} are needed.

Recently, influences of the spin-filter effect in ferromagnetic-semiconductors~\cite{rf:Meservey,rf:Moodera1,rf:Inomata} on the proximity effect~\cite{rf:Tokuyasu,rf:Cuevas,rf:Zhao,rf:Linder1,rf:Bergeret1,rf:Machon,rf:Giazotto13}, the Josephson effect~\cite{rf:Tanaka,rf:Fogelstrom,rf:Nishida,rf:Kawabata1,rf:Kawabata2,rf:Linder2,rf:Hikino1,rf:Liao,rf:Bergeret2,rf:Bergeret3,rf:Hikino2}, and macroscopic quantum phenomena~\cite{rf:Kawabata3,rf:Kawabata4,rf:Kawabata5,rf:Kawabata6} have been investigated theoretically.
Moreover, superconducting tunnel junctions with spin-filters have been also realized experimentally~\cite{rf:Senapati,rf:Golod,rf:Pal,rf:Muduli}
In this work we propose an novel electron-cooler based on clean N/spin-filter/S junctions [see Fig. 1(a)] and show that the cooling power is drastically enhanced due to the suppression of the Andreev reflection by the spin-filter effect as described in Fig. 1(b).
Preliminary result of this work has been reported in \cite{rf:Kawabata7}.
In this paper we will discuss about the theoretical derivation of the cooling power in more detail.

%%%%%%%%%%%
%%%%%%%%%%%
%%%%%%%%%%%
%%%%%%%%%%%
\begin{figure}[b]
\begin{center}
\includegraphics[width=8.1cm]{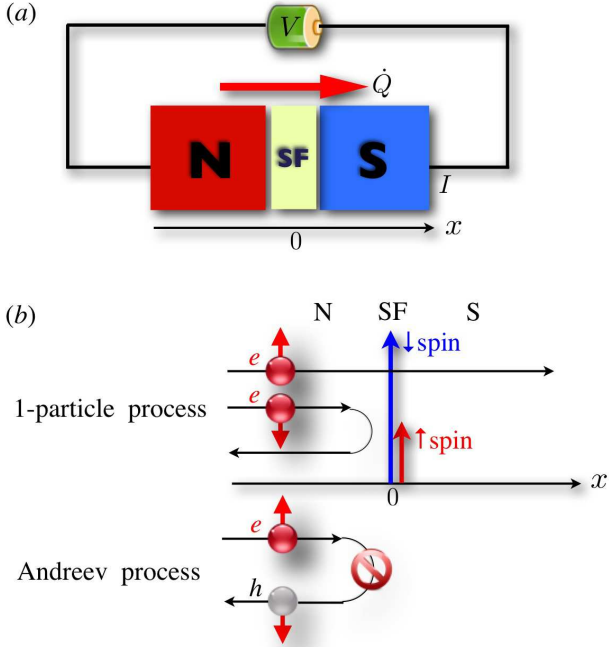}
\end{center}
\caption{(a) Schematic diagram of a normal-metal/spin-filter/superconductor (N/SF/S) cooler and (b) the delta-function model of a SF barrier.
In the SF interface ($x=0$), the transmission probability of electrons or holes for one spin-channel is much larger than the other one.
This allows the suppression of the Andreev reflection at the SF interface.
 }
\label{fig1}
\end{figure}
%%%%%%%%%%%
%%%%%%%%%%%
%%%%%%%%%%%
%%%%%%%%%%%

\section{Theory}

Let us first consider an one-dimensional ballistic N/SF/F junction as shown in Fig.~1(a). 
The spin-filtering barrier at $x=0$ can be described by a spin-dependent delta-function potential [see  Fig.~1(b)], $i. e.$, 
\begin{eqnarray}
V_\sigma (x) = \left( V + \rho_\sigma U \right) \delta(x),
\end{eqnarray} 
where $V$ is a spin-independent part of the potential, $U$ is the exchange-splitting, and $\rho_\sigma=+(-)1$ for up (down) spins~\cite{rf:Tanaka,rf:Moodera2}.

The spin-filtering property of the barrier is qualitatively characterized by the spin-filtering efficiency 
\begin{eqnarray}
P
=
\frac{\left| t_\uparrow - t_\downarrow \right|}{t_\uparrow+t_\downarrow}
\
,
\end{eqnarray} 
where 
\begin{eqnarray}
t_\sigma=\frac{1}{1+ (Z+ \rho_\sigma S)^2},
\end{eqnarray} 
is the transmission probability of the spin-filtering barrier for spin $\sigma$ with $m$ and $k_F$ being the mass of electrons and the Fermi wave number.
The normalized spin-independent and -dependent potential barrier-height are given by
\begin{eqnarray}
Z &\equiv& \frac{mV}{\hbar^2 k_F},
\\
S &\equiv& \frac{m U} {\hbar^2 k_F}
.
\end{eqnarray} 
For a perfect spin-filter with $t_\uparrow > 0$ and $t_\downarrow=0$, we get $P=1$.
On the other hand, we have $P=0$ for the conventional non-magnetic barrier with $U=0$ ($t_\uparrow =t_\downarrow$).

The system can be described by the Bogoliubov-de Gennes (BdG) equation~\cite{rf:Tanaka}:
\begin{align}
\left[
\begin{array}{cc}
H_0-\rho_\sigma U \delta (x) & \Delta(x) \\
\Delta^* (x)  & -H_0+\rho_\sigma U(x) \delta (x)
\end{array}
\right]
\Phi_\sigma (x)
\nonumber\\
=E
\Phi_\sigma (x)
,
\end{align} 
where $H_0$ is the spin-independent part of the single-particle Hamiltonian, $i. e.$,
\begin{align}
H_0= -\frac{\hbar^2 \nabla^2}{2 m}+V \delta(x)-\mu_F
,
\end{align} 
($\mu_F$ is the chemical potential), 
\begin{equation}
\Delta(x)=\Delta (T) e^{i \phi} \Theta(x)
\end{equation} 
is a pair potential [$\phi$ is the phase of the pair potential and $\Theta(x)$ is the Heaviside step function],
\begin{eqnarray}
\Phi_{\uparrow} (x)&=&
\left[
\begin{array}{c}
u_{\uparrow}(x) \\
v_{\downarrow}(x) 
\end{array}
\right]
,
\\
\Phi_{\downarrow} (x)&=&
\left[
\begin{array}{c}
u_{\downarrow}(x) \\
v_{\uparrow}(x) 
\end{array}
\right]
\end{eqnarray} 
are the eigenvectors, and the eigenenergy $E$ is measured from $\mu_F$.

The wave function in N ($x<0$) and S  ($x>0$) is given by 
\begin{eqnarray}
\Psi^N_\sigma (x)
&=&
\left(
\begin{array}{c}
1 \\
0
\end{array}
\right)
e^{i k^+ x}
+
\left(
\begin{array}{c}
1 \\
0
\end{array}
\right)
e^{-i k^+ x}
r_\sigma^{ee}
\nonumber 
\\
&+&
\left(
\begin{array}{c}
0 \\
1
\end{array}
\right)
e^{i k^- x}
r_\sigma^{he}
,
\end{eqnarray} 
\begin{align}
\Psi^S_\sigma (x)
=
\left(
\begin{array}{c}
u_0 \\
v_0 e^{-i \phi}
\end{array}
\right)
e^{i q^+ x}
t_\sigma^{ee}
+
\left(
\begin{array}{c}
v_0 e^{i \phi}\\
u_0 
\end{array}
\right)
e^{-i q^- x}
t_\sigma^{he}
,
\end{align} 
where 
\begin{eqnarray}
u_0&= &\sqrt{\frac{1}{2} \left(  1+ \frac{\Omega}{E} \right)}
,
\\
v_0&= &\sqrt{\frac{1}{2} \left(  1- \frac{\Omega}{E} \right)}
,
\\
k^\pm&= &k_F \sqrt{1\pm \frac{E}{\mu_F}}
,
\\
q^\pm&=& k_F \sqrt{1\pm \frac{\Omega}{\mu_F}}
,
\end{eqnarray} 
with 
\begin{eqnarray}
\Omega=i \sqrt{\Delta(T)^2-E^2}
.
\end{eqnarray} 
The normal reflection coefficient $r_\sigma^{ee}$ and the Andreev reflection coefficient $r_\sigma^{he}$ can be obtained by solving the BdG equation with two boundary conditions at the spin-filtering barrier ($x=0$):
\begin{align}
\Psi^N_\sigma (0)=\Psi^S_\sigma (0)
,
\end{align} 
\begin{eqnarray}
&-& \frac{\hbar^2}{2 m} 
\left(
 \left. \frac{d}{d x} \Psi_\sigma^S (x) \right|_{x=0} 
-
 \left. \frac{d}{d x} \Psi_\sigma^N (x) \right|_{x=0} 
\right)
\nonumber
\\
&+&
\left[
\begin{array}{cc}
V-\rho_\sigma U &0 \\
0 & V+\rho_\sigma U
\end{array}
\right]
\Psi_\sigma^S (0)
=0
.
\end{eqnarray} 
By assuming  
\begin{eqnarray}
k^\pm \approx q^\pm \approx k_F
\end{eqnarray} 
based on the fact that $E \sim \Delta(T) \ll \mu_F$, we can analytically obtain $r_\sigma^{ee}$ and  $r_\sigma^{he}$ as follows:
\begin{eqnarray}
r_\sigma^{ee}(E)
&=&
-
\frac{2 
\left[  
 i \left( \Omega Z - |E| \rho_\sigma S \right)
 + \Omega (Z^2 - S^2)
\right]}{  \left( |E| + \Omega \right) \left( 1 - 2 i \rho_\sigma S\right)+ 2 \Omega  (Z^2 - S^2)}
,
\nonumber
\\
\\
r_\sigma^{he}(E)
&=&
\ \ \ 
\frac{
\Delta (T) e^{- i \phi}
}{  \left( |E| + \Omega \right) \left( 1 - 2 i \rho_\sigma S\right)+ 2 \Omega  (Z^2 - S^2)}.
\nonumber
\\
\end{eqnarray} 
In the following calculations, we have determined the temperature $T$ dependence of the superconducting gap $\Delta(T)$ by solving the BCS gap equation numerically.

In order to check the suppression of the Andreev reflection by the spin-filter effect, firstly we study the spin-dependent electron transport of the junction.
The voltage $V$ dependence of the differential conductance $G$ of the system can be calculated from the Blonder-Tinkham-Klapwijk formula~\cite{rf:Blonder},
\begin{align}
G (V) =\frac{e^2}{h} 
\sum_{\sigma=\uparrow,\downarrow}
\left[
  1- B_\sigma(E=eV) + A_\sigma(E=eV)
\right] 
,
\end{align} 
where 
\begin{align}
B_\sigma(E)  \equiv \left| r_\sigma^{ee}(E)  \right|^2 
,
\nonumber\\
A_\sigma(E)  \equiv \left| r_\sigma^{he} (E)\right|^2 
.
\end{align} 
In Fig.~2 we plot the spin-filtering efficiency $P$ dependence of the conductance $G(V)/G_N$ vs $eV/\Delta_0$ for a junction with (a) the transparent  ($t_\uparrow=1.0$) and (b) the tunneling barrier ($t_\uparrow=0.1$) at zero temperature, where
\begin{align}
G_N =\frac{2 e^2}{h} 
\frac{1+Z^2+S^2}{\left( Z^2-S^2+1\right)^2 + 4 S^2}
,
\end{align} 
stands for the conductance of an N/SF/N junction and $\Delta_0=\Delta(T=0$K).
If $P$ is increased, the sub-gap conductance for $|eV| \le \Delta_0$  is largely reduced~\cite{rf:Kawabata7,rf:Kashiwaya}.
It is important to note that for the case of the perfect spin-filter ($P=1$), the Andreev reflection is completely inhibited, indicating that the spin-filter would suppress the unwanted Andreev Joule heating.
%%%%%%%%%%%
%%%%%%%%%%%
%%%%%%%%%%%
%%%%%%%%%%%
\begin{figure}[t]
\begin{center}
\includegraphics[width=8.0cm]{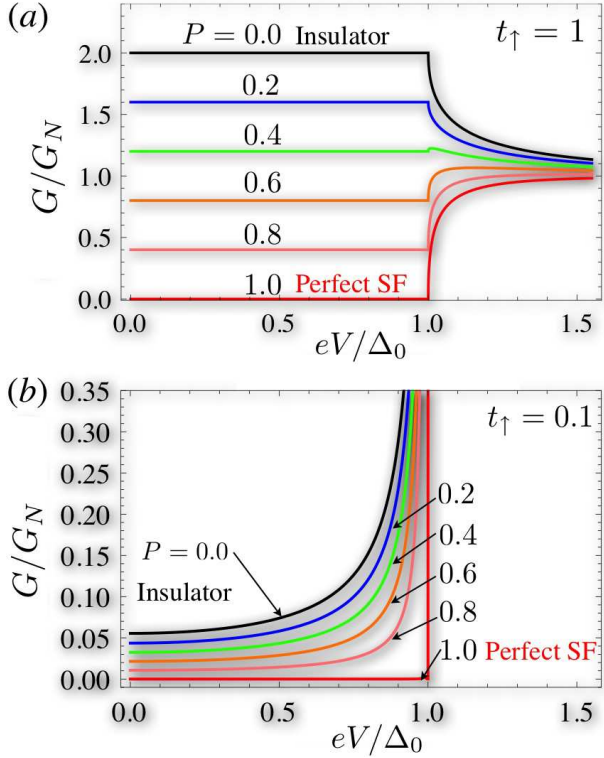}
\end{center}
\caption{The conductance $G$ vs the bias voltage $V$  of a normal-metal/spin-filter/superconductor (N/SF/S) cooler at $T=0$K for (a) transparent  ($t_\uparrow=1$) and (b) tunneling barrier ($t_\uparrow=0.1$).
$G_N$ is the conductance of a N/SF/N junction, $\Delta_0$ is the superconducting gap at $T=0$K, and $P$ is the spin-filtering efficiency, respectively. $P=0.0 (1.0)$ is corresponding to a nonmagnetic-insulating (a perfect SF) interlayer.
By increasing $P$, the sub-gap conductance is reduced considerably due to the suppression of the Andreev reflection.
 }
\label{fig2}
\end{figure}
%%%%%%%%%%%
%%%%%%%%%%%
%%%%%%%%%%%
%%%%%%%%%%%

%%%%%%%%%%%
%%%%%%%%%%%
%%%%%%%%%%%
%%%%%%%%%%%
\begin{figure}[t]
\begin{center}
\includegraphics[width=8.0cm]{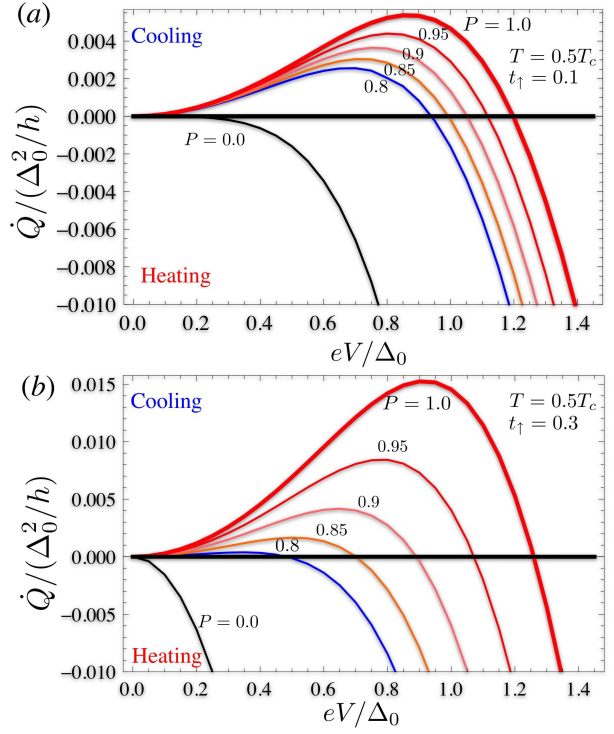}
\end{center}
\caption{The cooling power ${\dot Q}$ vs bias voltage $V$  of an N/SF/S cooler with (a) $t_\uparrow=0.1 $ and (b) 0.3 at $T=0.5T_c$ for several spin polarizations $P$.
 }
\label{fig3}
\end{figure}
%%%%%%%%%%%
%%%%%%%%%%%
%%%%%%%%%%%
%%%%%%%%%%%

In order to see the benefit of the spin-filtering barrier on the electron cooling, we numerically calculate the cooling power by 
using the Bardas and Averin formula~\cite{rf:Giazotto06,rf:Bardas},
\begin{eqnarray}
\dot{Q} (V)
&=&\frac{2 e}{h} 
\sum_{\sigma=\uparrow,\downarrow}
\int_{-\infty}^\infty d E 
\left[ E \left\{ 1- B_\sigma(E) - A_\sigma(E) \right\}
\right.
\nonumber
\\
&-& 
\left.
eV \left\{ 1-B_\sigma(E) +A_\sigma(E) \right\}
\right] 
\nonumber
\\
&\times& 
\left[  f(E-eV) -f(E)\right]
,
\end{eqnarray} 
where $f(E)$ stands for the Fermi-Dirac distribution function. 
In the case of $\dot{Q}>0$, we can realize cooling of N.

The cooling power $\dot{Q}$ vs the bias voltage $V$ for (a) $t_\uparrow=0.1$ and (b) $t_\uparrow=0.3$ is shown in Fig. 3.
In the calculation we have set that $T=0.5 T_c$, where $T_c$ is the superconducting transition temperature.
As will be discussed later, the maximal cooling power can be realized for $T \approx 0.5 T_c$ see Fig.~4(a).
If we increase $P$, the cooling power $\dot{Q}$ is enhanced drastically.
These peculiar results can be attributed to the suppression of the Andreev reflection and equivalently the undesirable Andreev Joule heating.
This means that the spin-filter effect dramatically boosts the cooling power $\dot{Q}$ in comparison with conventional N/I/S coolers.

%%%%%%%%%%%
%%%%%%%%%%%
%%%%%%%%%%%
%%%%%%%%%%%
\begin{figure}[t]
\begin{center}
\includegraphics[width=8.0cm]{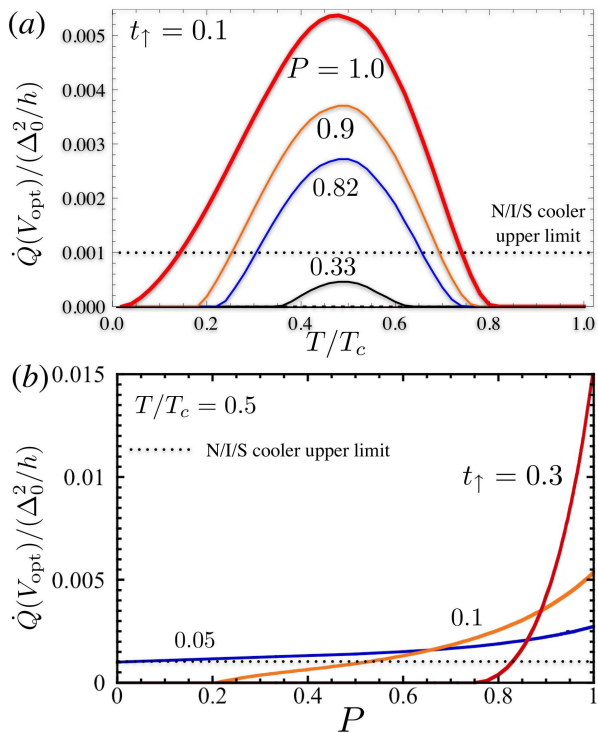}
\end{center}
\caption{
The optimal cooling power $\dot{Q}(V_\mathrm{opt})$ of a N/SF/S cooler as a function of (a) temperature $T$ and (b) the spin-filtering efficiency $P$. 
The dotted line is the theoretical upper limit of the optimal cooling power $\dot{Q}^\mathrm{max} (V_\mathrm{opt})  \approx 0.001 \Delta_0^2 / h$  for conventional N/I/S coolers, which can be achieved in the case of $T/T_c \approx 0.5$ and $t_\uparrow=t_\downarrow \approx 0.05$.
 }
\label{fig4}
\end{figure}
%%%%%%%%%%%
%%%%%%%%%%%
%%%%%%%%%%%
%%%%%%%%%%%

Next let us discuss about the optimization of the cooling power in terms of temperature $T$ as well as the spin-filtering efficiency $P$ to design the high-performance cooler.
In Fig.~4 we plot the cooling power $\dot{Q}$ as a function of temperature $T$ at the optimal bias voltage $V=V_\mathrm{opt}$ in which $\dot{Q}$ is maximized as function of $V$.
The theoretical upper-limit of the cooling power for conventional N/I/S coolers [$\dot{Q}(V_\mathrm{opt}) \approx 0.001 (\Delta_0^2/ h)$] is realized for $t_\uparrow =t_\downarrow \approx 0.05$ and $T/T_c  \approx 0.5$ (see the dotted lines in Fig.~4).
As clearly seen from Fig.~4(a), $\dot{Q}(V_\mathrm{opt})$ is maximized around $T \approx 0.5 T_c$, decreasing at both higher and lower temperatures.
From the view point of practical applications, it is remarkable that if we increase $P$, both the maximum value of $\dot{Q}$ and the minimum temperature in which $\dot{Q}(V_\mathrm{opt}) \ge 0 $ are largely improved.
Especially latter property means that we can achieve much lower temperature than the case of conventional N/I/S coolers by the help of the spin-filter effect.

In order to realize high-performance refrigerator, the large cooling power  $\dot{Q}$ is needed.
We also plot  the spin-filtering efficiency $P$ dependence of the optimal cooling-power $\dot{Q}(V_\mathrm{opt})$ for different values of $t_\uparrow$ in Fig.~4(b).
The maximum cooling-power $\dot{Q}(V_\mathrm{opt})$ for N/SF/S junctions can be achieved in the case of the perfect spin-filter ($P=1$) because of the complete suppression of the Andreev reflection.
It is important to note that even in the small $P$ value ($P \ll 1$), $\dot{Q}(V_\mathrm{opt})$ overcomes the theoretical upper-limit for conventional N/I/S coolers.
More notably, for the case of $t_\uparrow=0.3$, $\dot{Q}(V_\mathrm{opt})$ can be a factor of 15 larger than the theoretical upper-limit for N/I/S coolers.

Based on above results, we next discuss about the advantage of SF-based coolers over FM-based ones (N/FM/S junctions)~\cite{rf:Giazotto02,rf:Ozaeta}.
In order to realize positive cooling-power for FM-based coolers, it was found that considerably high spin-polarization $P>0.94$ is needed.
In this sense we have to use exotic and recalcitrant FMs, like half-metals~\cite{rf:Wolf} in the FM interlayer.
On the other hand, in N/SF/S cooler, much smaller value of $P$ is enough for realizing the high performance cooler.
This means that large number of SF materials, $e. g.,$ Eu chalcogenides~\cite{rf:Moodera1}, rare earth nitrides, spinel ferrites~\cite{rf:Luders,rf:Takahashi,rf:Matzen,rf:Caffrey,rf:Nozaki}, and manganites~\cite{rf:Golod,rf:Satapathy,rf:Harada1,rf:Liu,rf:Harada2} can be used for solid-state coolers.
More importantly such junctions with large $P$ have been already realized in an EuS/Al ($P \sim 0.9$)~\cite{rf:Moodera1}, EuSe/Al($P \sim 1$)~\cite{rf:Moodera2}, and GdN/NbN junction ($P \sim 0.8$)~\cite{rf:Senapati}. 
Therefore we can conclude that SF-based cooler is much more practical than the FM based one.
This is a crucial advantage of the SF-based cooler.

It is important to note that in spin-filter coolers with large $P$, one of the spins (e.g., up-spin electrons) with $E>\Delta(T)$ can tunnel through the SF barrier, but opposite spins (e.g., down-spin electrons) with $E>\Delta(T)$ can not be escaped from N to S.
This means that the only the up-spin electrons can contribute to the cooling.
By using a S/SF1/N/SF2/S structure in which the magnetization direction of SF1 and SF2 layers is antiparallel, it is possible to effectively cool down both up- and down-spin quasiparticles in N.

\section{Summary}
To summarize, we have proposed a novel electron-cooler based on ballistic N/SF/S junctions.
We found that the cooling power $\dot{Q}$ is higher than the theoretical upper-limit of $\dot{Q}$ for conventional N/I/S coolers, which results form the suppression of the Andreev Joule heating.
Our results open up a way to make efficient solid-based refrigerators for cooling several useful and practical devices, such as superconducting X-ray detectors, single-photon detectors, magnetic sensors, NEMSs, and qubits.

\section*{Acknowledgements}
We would like to S. Nakamura for useful discussions.
This work was supported by the Topological Quantum Phenomena (No.23103520) KAKENHI on Innovative Areas, a Grant-in-Aid for Scientific Research
(No. 25286046) from MEXT of Japan, the JSPS Institutional Program for Young Researcher Overseas Visits, the European Union Seventh Framework Programme (FP7/2007-2013) under grant agreement "INFERNOS" No. 308850, the Spanish Ministry of Economy and Competitiveness under Project FIS2011-28851-C02-02, and the CSIC and the European Social Fund under JAE-Predoc program.

\end{document}